\documentclass[letter,longauth,traditabstract]{aa} 
\usepackage{graphicx}
\usepackage{txfonts}
\begin{document}
   \title{Enhanced dust heating in the bulges of early-type spiral
galaxies\thanks{Herschel is an ESA space observatory with science instruments
provided by the European-led Principal Investigator consortia and with important
participation from NASA.}}

   \titlerunning{FIR to submm. disk and central SED comparison}

   \author{C.~W.\ Engelbracht\inst{1}
           \and L.~K.\ Hunt\inst{2}
           \and R.~A.\ Skibba\inst{1}
           \and J.~L.\ Hinz\inst{1}
           \and D.\ Calzetti\inst{3}
           \and K.~D.\ Gordon\inst{4}
           \and H.\ Roussel\inst{5}
           \and A.~F.\ Crocker\inst{3}
           \and K.~A.\ Misselt\inst{1}
           \and A.~D.\ Bolatto\inst{6}
           \and R.~C.\ Kennicutt\inst{7}
           \and P.~N.\ Appleton\inst{8}
           \and L.\ Armus\inst{9}
           \and P.\ Beir\~{a}o\inst{9}
           \and B.~R.\ Brandl\inst{10}
           \and K.~V.\ Croxall\inst{11}
           \and D.~A.\ Dale\inst{12}
           \and B.~T.\ Draine\inst{13}
           \and G.\ Dumas\inst{14}
           \and A.\ Gil de Paz\inst{15}
           \and B.\ Groves\inst{10}
           \and C.-N.\ Hao\inst{16}
           \and B.~D.\ Johnson\inst{7}
           \and J.\ Koda\inst{17}
           \and O.\ Krause\inst{14}
           \and A.~K.\ Leroy\inst{18} \thanks{Hubble Fellow}
           \and S.~E.\ Meidt\inst{14}
           \and E.~J.\ Murphy\inst{9}
           \and N.\ Rahman\inst{19}
           \and H.-W.\ Rix\inst{14}
           \and K.~M.\ Sandstrom\inst{14}
           \and M.\ Sauvage\inst{20}
           \and E.\ Schinnerer\inst{14}
           \and J.-D.~T.\ Smith\inst{11}
           \and S.\ Srinivasan\inst{5}
           \and L.\ Vigroux\inst{5}
           \and F.\ Walter\inst{14}
           \and B.~E.\ Warren\inst{21}
           \and C.~D.\ Wilson\inst{19}
           \and M.~G.\ Wolfire\inst{6}
           \and S.\ Zibetti\inst{14}
          }

   \institute{Steward Observatory, University of Arizona, Tucson, AZ 85721,
              USA \\
              \email{cengelbracht@as.arizona.edu}
         \and INAF - Osservatorio Astrofisico di Arcetri, Largo E. Fermi 5,
              50125 Firenze, Italy
         \and Department of Astronomy, University of Massachusetts, Amherst,
              MA 01003, USA
         \and Space Telescope Science Institute, 3700 San Martin Drive,
              Baltimore, MD 21218, USA
         \and Institut d'Astrophysique de Paris, UMR7095 CNRS,
              Universit\'e Pierre \& Marie Curie, 98 bis Boulevard Arago,
              75014 Paris, France 
         \and Department of Astronomy, University of Maryland, College Park,
              MD 20742, USA
         \and Institute of Astronomy, University of Cambridge, Madingley Road,
              Cambridge CB3 0HA, UK
         \and NASA Herschel Science Center, IPAC, California Institute of
              Technology, Pasadena, CA 91125, USA
         \and Spitzer Science Center, California Institute of Technology, MC
              314-6, Pasadena, CA 91125, USA
         \and Leiden Observatory, Leiden University, P.O. Box 9513, 2300 RA
              Leiden, The Netherlands
         \and Department of Physics and Astronomy, University
              of Toledo, 2801 West Bancroft Street, Toledo, OH 43606, USA
         \and Department of Physics \& Astronomy, University of Wyoming,
              Laramie, WY 82071, USA
         \and Department of Astrophysical Sciences, Princeton University,
              Princeton, NJ 08544, USA
         \and Max-Planck-Institut f\"{u}r Astronomie, K\"{o}nigstuhl 17, 69117
              Heidelberg, Germany
         \and Departamento de Astrofisica, Facultad de Ciencias Fisicas,
              Universidad Complutense Madrid, Ciudad Universitaria, Madrid, E-28040, Spain
         \and Tianjin Astrophysics Center, Tianjin Normal University, 
              Tianjin 300387, China
         \and Department of Physics and Astronomy, SUNY Stony Brook, Stony
              Brook, NY 11794-3800, USA
         \and National Radio Astronomy Observatory, 520 Edgemont Road, 
              Charlottesville, VA 22903, USA
         \and Department of Physics \& Astronomy, McMaster University, Hamilton,
              Ontario L8S 4M1, Canada
         \and CEA/DSM/Irfu/Service d'Astrophysique, UMR AIM, CE Saclay,
              91191 Gif sur Yvette Cedex, France
         \and ICRAR, M468, University of Western Australia, 35 Stirling Hwy,
              Crawley, WA, 6009, Australia 
             }

  \abstract
   {Stellar density and bar strength should affect the temperatures of the
cool (T~$\sim 20-30$~K) dust component in the inner regions of galaxies, which
implies that the ratio of temperatures in the circumnuclear regions to the
disk should depend on Hubble type.  We investigate the differences between
cool dust temperatures in the central 3~kpc and disk of 13 nearby galaxies by
fitting models to measurements between 70 and 500~$\mu$m.  We attempt to
quantify temperature trends in nearby disk galaxies, with archival data from
\emph{Spitzer}/MIPS and new observations with \emph{Herschel}/SPIRE, which
were acquired during the first phases of the \emph{Herschel} observations for
the KINGFISH (key insights in nearby galaxies: a far-infrared survey with
\emph{Herschel}) sample.  We fit single-temperature modified blackbodies to
far-infrared and submillimeter measurements of the central and disk regions of
galaxies to determine the temperature of the component(s) emitting at those
wavelengths.  We present the ratio of central-region-to-disk-temperatures of
the cool dust component of 13 nearby galaxies as a function of morphological
type.  We find a significant temperature gradient in the cool dust component
in all galaxies, with a mean center-to-disk temperature ratio of $1.15 \pm
0.03$.  The cool dust temperatures in the central $\sim3$~kpc of nearby
galaxies are 23 ($\pm3$)\% hotter for morphological types earlier than Sc,
and only 9 ($\pm3$)\% hotter for later types.  The temperature ratio
is also correlated with bar strength, with only strongly barred galaxies
having a ratio over 1.2.  The strong radiation field in the high
stellar density of a galactic bulge tends to heat the cool dust component to
higher temperatures, at least in early-type spirals with relatively large
bulges, especially when paired with a strong bar.}


   \maketitle
%

\section{Introduction}

The infrared emission from galaxies contains roughly half of the entire energy
budget in the Universe (e.g., Hauser \& Dwek \cite{hauser01}). In addition to
providing information on the amount of attenuation suffered by the stellar
light, the infrared emission provides clues to important physical quantities,
such as the metal, dust, and cold gas content of galaxies (e.g., Draine et
al.\ \cite{draine07a}, Bernard et al.\ \cite{bernard08}).  Infrared spectral
energy distributions (SEDs), especially those extending into the
submillimeter regime, can be used to measure the dust mass and temperature
in galaxies (e.g., Dunne et al. \cite{dunne00,dunne01}; Seaquist et al.\
\cite{seaquist04}; Vlahakis et al.\ \cite{vlahakis05}; Draine et al.\
\cite{draine07a}; Willmer et al.\ \cite{willmer09}; Liu et al.\ \cite{liu10}).

These temperature measurements have shown that the dust is warmer in the
centers of galaxies than in the outskirts (e.g., Alton et al.\ \cite{alton98};
Radovich et al.\ \cite{radovich01}; Melo et al.\ \cite{melo02}; Dupac et al.\
\cite{dupac03}).  Cool dust at roughly the same temperature in spiral disks is
detected {\it globally} at longer wavelengths (850~$\mu$m and 1.3~mm;
Siebenmorgen et al.\ \cite{siebenmorgen99}; Dunne et al.\
\cite{dunne00,dunne01}; Vlahakis et al.\ \cite{vlahakis05}), but there is also
some evidence of a warmer temperature component associated with the central
regions.  Warmer dust temperatures tend to be associated with significant
star-formation activity, the resulting intense interstellar radiation
field (Stevens et al.\ \cite{stevens05}) and the earlier Hubble type
(Bendo et al.\ \cite{bendo03}).

Here we use the infrared SEDs, from 70 to 500~$\mu$m (a range which should be
dominated by emission from grains in thermal equilibrium with the radiation
field, e.g., Popescu et al.\ \cite{popescu00}, Engelbracht et al.\
\cite{engelbracht08}), of a local sample of 13 galaxies spanning the range of
spiral galaxies in the Hubble sequence, to derive the temperature of the cool
(T$ \sim 20-30$~K) dust component of the central region and the disks
separately, and investigate differences in the dust heating in the two
regions.  To achieve this goal, we use the 250, 350, and 500~$\mu$m
\emph{Herschel}/SPIRE (Spectral and Photometric Imaging REceiver; Griffin et
al.\ \cite{griffin10}) images of the galaxies combined with the
\emph{Spitzer}/MIPS (Multiband Imaging Photometer for \emph{Spitzer}; Rieke
et~al.\ \cite{rieke04}) 70 and 160~$\mu$m images.  Eventually we will use PACS
(Photodetector Array Camera and Spectrometer; Poglitsch et al.\
\cite{poglitsch10}) imaging for this study, but at the time of this writing,
the data were not yet available.  These galaxies have little nuclear activity
that might heat the dust, with only NGC~1097 having a strongly active nucleus,
so this is the first study to cleanly separate the far-infrared properties of
central and disk regions in a sample of normal galaxies. 

Until recently, little work has been done to dissect the dust emission in
galaxies into sub-galactic components, owing to the general paucity of
infrared images with the required angular resolution and to poor
long-wavelength sensitivity.  Some recent work includes a study of the galaxy
pair NGC1512/1510 (a target also discussed in this paper), which finds that
the dust temperature in the central region of NGC1512 is slightly higher than
in the disk and that there is a significantly higher fraction of warm dust, in
agreement with the center of NGC1512 being a starburst (Liu et al.\
\cite{liu10}).  Work by Mu\~{n}oz-Mateos et al.\ (\cite{munoz-mateos09})
examines radial trends in dust properties in a number of nearby galaxies.

The \emph{Herschel} Space Telescope promises to yield a breakthrough in the
study of subgalactic components in galaxies.  This paper is the first
investigation to leverage the longest wavelength \emph{Herschel} data available
for the KINGFISH (key insights on nearby galaxies: a far-infrared survey with
\emph{Herschel}; this program is largely derived from SINGS, the
\emph{Spitzer} infrared nearby galaxy survey by Kennicutt et al.\
\cite{kennicutt03}) sample of nearby galaxies, which will eventually total 61.
Companion papers from the science demonstration phase for this program
showcase the shorter wavelength imaging (Sandstrom et al.\
\cite{sandstrom10}) and the spectroscopic data (Beir\~{a}o et al.\
\cite{beirao}).

Here we present new SPIRE images acquired during the first few months of
\emph{Herschel} operations, in the context of the KINGFISH Open-Time Key
Project.  We divide each galaxy into two spatially-resolved zones: the
circumnuclear region and the surrounding disk.  Then we compare central
temperatures with those for the disk.


\section{Observations and data reduction}

We observed 33 nearby galaxies with the SPIRE instrument on \emph{Herschel}
(Pilbratt et al.\ \cite{pilbratt10}) in the scan map mode as part of the
KINGFISH program over the period of November 2009 to January 2010.  They were
observed in the scan mode out to 1.5 optical radii, to depths of (3.2, 2.5,
and 2.9)~mJy/beam at (250, 350, and 500)~$\mu$m.  They were reduced using the
standard calibration products and algorithms available in HIPE (the
\emph{Herschel} Interactive Processing Environment; Ott \cite{ott10}), version
1.2.5 or 2.0.0 (whichever version was the latest available when the target was
observed, since the SPIRE calibration is not sensitive to this range of HIPE
versions).  We modified the offset-subtraction algorithm in the pipeline, by
requiring it to measure the offset only outside bright objects in the field
(usually the galaxy that we targeted)---this procedure reduced a small
negative offset (6\% in the first galaxy we observed, NGC~4559) in the
background in the same rows and columns as the source to an undetectable
level.  Otherwise, the data reduction was performed using the default
pipeline.

\section{Analysis}

\begin{figure*}
\centering
\includegraphics[width=\textwidth]{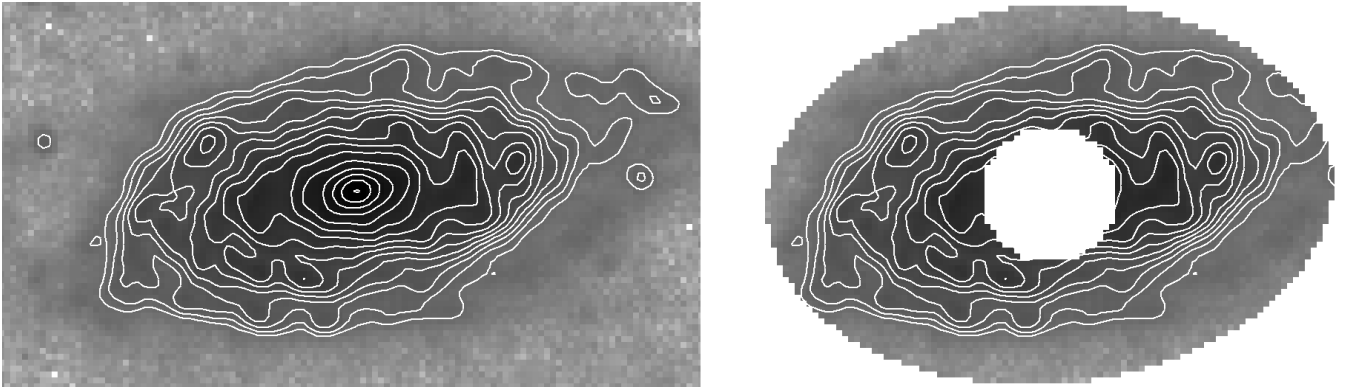}
\caption{Example of the masking approach used, shown here at 250~$\mu$m on a
cropped image of NGC~5055.  The image on the left shows the full, reduced
image, while to the image on the right we have applied a mask (the blank,
white regions) to all data but the disk.  Both images have been scaled
logarithmically, from -0.01 to 5 Jy/beam.  The contour levels are scaled
logarithmically from 0.1 to 5 Jy/beam. Each panel is 6\arcmin\
across.\label{fig:masking}}
\end{figure*}

Of the 33 galaxies observed, we selected 13 which had both a large extent
(i.e., greater than several 160/500~$\mu$m beam diameters, or $\sim2$\arcmin)
and a bright disk (brighter than $\sim0.5$~Jy at 500~$\mu$m, excluding the
nucleus).  These galaxies are listed in Table~\ref{tab:sample}.  For the
wavelengths with significantly smaller beams (i.e., 70, 250, and 350~$\mu$m)
than the largest used here, we convolved the data to the SPIRE 500~$\mu$m
resolution using the kernels described by Gordon et al.\ (\cite{gordon08}) and
updated by those authors for \emph{Herschel}.  For each galaxy, we computed
masks that isolated the disk and central regions.  The outer mask is an
ellipse sized to encompass the disk where it contrasts strongly with the
background, at 1.2, 6.6, 2.7, 1.4, and 0.8~MJy/sr at 70, 160, 250, 350, and
500~$\mu$m (where the SPIRE data have been converted to MJy/sr as described
below), respectively.  The inner mask is a circle centered on the galaxy peak,
with a minimum size chosen to be as small as possible (i.e., set by the MIPS
160~$\mu$m beam) and with a radius inversely proportional to the distance to
the galaxy.  Thus, a similar physical region of $\sim3$~kpc diameter (in which
one would expect to find structures like a bulge, bar, ring, and/or inner
disk, e.g., see Erwin \& Sparke \cite{erwin02}, Athanassoula \& Martinet
\cite{athanassoula80}, and Knapen \cite{knapen05}) is sampled in the central
region of each galaxy.  Our results are not sensitive to a modest change in
the aperture size, for example, similar results are achieved with a fixed
central aperture size of 40\arcsec\ for each galaxy.  A sample mask is shown
in Fig.~\ref{fig:masking}.

\begin{table}
\begin{minipage}{\columnwidth}
\caption[]{Galaxy sample and computed temperatures}
\label{tab:sample}
\centering
\renewcommand{\footnoterule}{}  
\tiny
\begin{tabular}{rrrrrr}
\hline\hline
Name & Hubble & D\footnote{As listed in NED: The NASA/IPAC
Extragalactic Database (NED) is operated by the Jet Propulsion Laboratory,
California Institute of Technology, under contract with the National
Aeronautics and Space Administration.} & Spectral\footnote{from NED or Ho
et al.\ \cite{ho97}}
& T$_{\rm disk}$\footnote{includes uncertainty in the photometric
calibration} & T$_{\rm center}$$^c$ \\
& type & (Mpc) & type & (K) & (K) \\
\hline
NGC0628 & SAc & 7.3 & HII? & 22.7$\pm$0.4 & 24.2$\pm$0.5 \\
NGC0925 & SABd & 9.0 & HII & 22.3$\pm$0.4 & 23.9$\pm$0.5 \\
NGC1097 & SBb & 19.1 & Sy1 & 23.4$\pm$0.5 & 28.7$\pm$0.7 \\
NGC1291 & SB0/a & 10.4 & \ldots & 18.3$\pm$0.3 & 27.6$\pm$0.6 \\
NGC1512 & SBa & 14.4 & HII & 21.3$\pm$0.4 & 27.3$\pm$0.6 \\
NGC3621 & SAd & 6.9 & \ldots & 21.9$\pm$0.4 & 25.6$\pm$0.6 \\
NGC4559 & SABcd & 8.5 & HII & 22.7$\pm$0.4 & 24.2$\pm$0.5 \\
NGC4594 & SAa & 9.4 & Sy1.9 & 21.3$\pm$0.4 & 22.9$\pm$0.4 \\
NGC4631 & SBd & 7.6 & HII & 24.5$\pm$0.5 & 27.7$\pm$0.6 \\
NGC5055 & SAbc & 10.2 & Transition2 & 22.0$\pm$0.4 & 25.3$\pm$0.5 \\
NGC5457 & SABcd & 7.1 & HII & 23.4$\pm$0.5 & 23.9$\pm$0.5 \\
NGC6946 & SABcd & 6.8 & HII & 24.5$\pm$0.5 & 26.6$\pm$0.6 \\
NGC7331 & SAb & 14.9 & Transition2 & 23.2$\pm$0.4 & 26.2$\pm$0.6 \\
\end{tabular}
\end{minipage}
\end{table}

We measured flux densities in these apertures for each of the 3
\emph{Herschel}/SPIRE bands, as well as the \emph{Spitzer}/MIPS 70 and
160~$\mu$m bands.  The \emph{Herschel}/SPIRE data were calibrated in units of
Jy/beam, which we converted to Jy by assuming beam sizes of 371, 720, and 1543
square arcseconds in the 250, 350, and 500~$\mu$m bands, respectively
(Swinyard et al.\ \cite{swinyard10}).  We assign uncertainties as the
quadratic sum of the data-dependent scatter and (by far the larger term) the
uncertainty in the absolute calibration, taken to be 5\% and 12\% for
\emph{Spitzer}/MIPS at 70 and 160~$\mu$m (Gordon et al.\ \cite{gordon07} and
Stansberry et al.\ \cite{stansberry07}, respectively) and 16\% (the quadratic
sum of the 15\% absolute calibration uncertainty from Swinyard et al.\
\cite{swinyard10} and a 2\% uncertainty in the size of the beam) for SPIRE
data.

\begin{figure}
\centering
\includegraphics[width=0.49\textwidth]{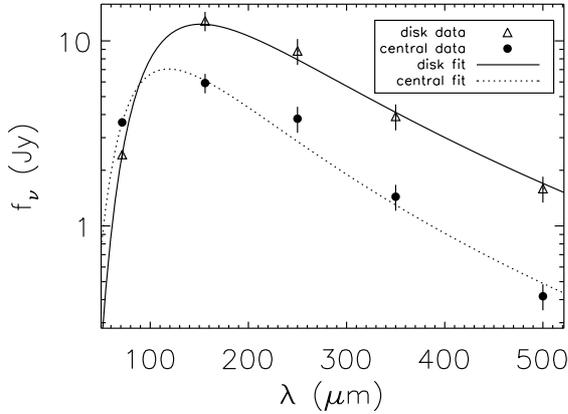}
\caption{Far-infrared SEDs of the disk and central region of NGC~1512, each
fit by a blackbody function modified by a $\lambda^{-1.5}$
emissivity.\label{fig:fit}}
\end{figure}

To each set of data, we fit a blackbody with a frequency-dependent emissivity
of 1.5.  Our conclusions do not change if values of 1 or 2 are used,
although the temperatures increase or decrease, respectively.  The
uncertainties in the temperatures were computed via a Monte Carlo simulation,
in which we performed 10000 trials where the photometric measurements
were allowed to vary in a normal distribution with a standard deviation
indicated by the photometric uncertainty.  An example of fits and data points
is shown in Fig.~\ref{fig:fit}, where we can see that the 250~$\mu$m flux
density is underpredicted by this simple model.  This happens frequently in
our dataset, and may be a sign that multiple temperature components are
present in the regions we measured.  The values of the \emph{Herschel}/SPIRE
uncertainty indicated by the uncertainty in the absolute calibration were
allowed to vary together; i.e., the calibration uncertainty was assumed to be
correlated among the three bands.  In contrast, we allowed the MIPS
uncertainties to vary independently; while not being strictly independent (see
Stansberry et al.\ \cite{stansberry07}), the 160~$\mu$m calibration
observations, in particular, are dominated by observational scatter and can be
treated independently of the 70~$\mu$m band.  The temperature of each
component is taken to be the temperature of the best-fit model, with an
uncertainty determined by the Monte Carlo simulation.

\section{Results}

\begin{figure}
\centering
\includegraphics[width=0.49\textwidth]{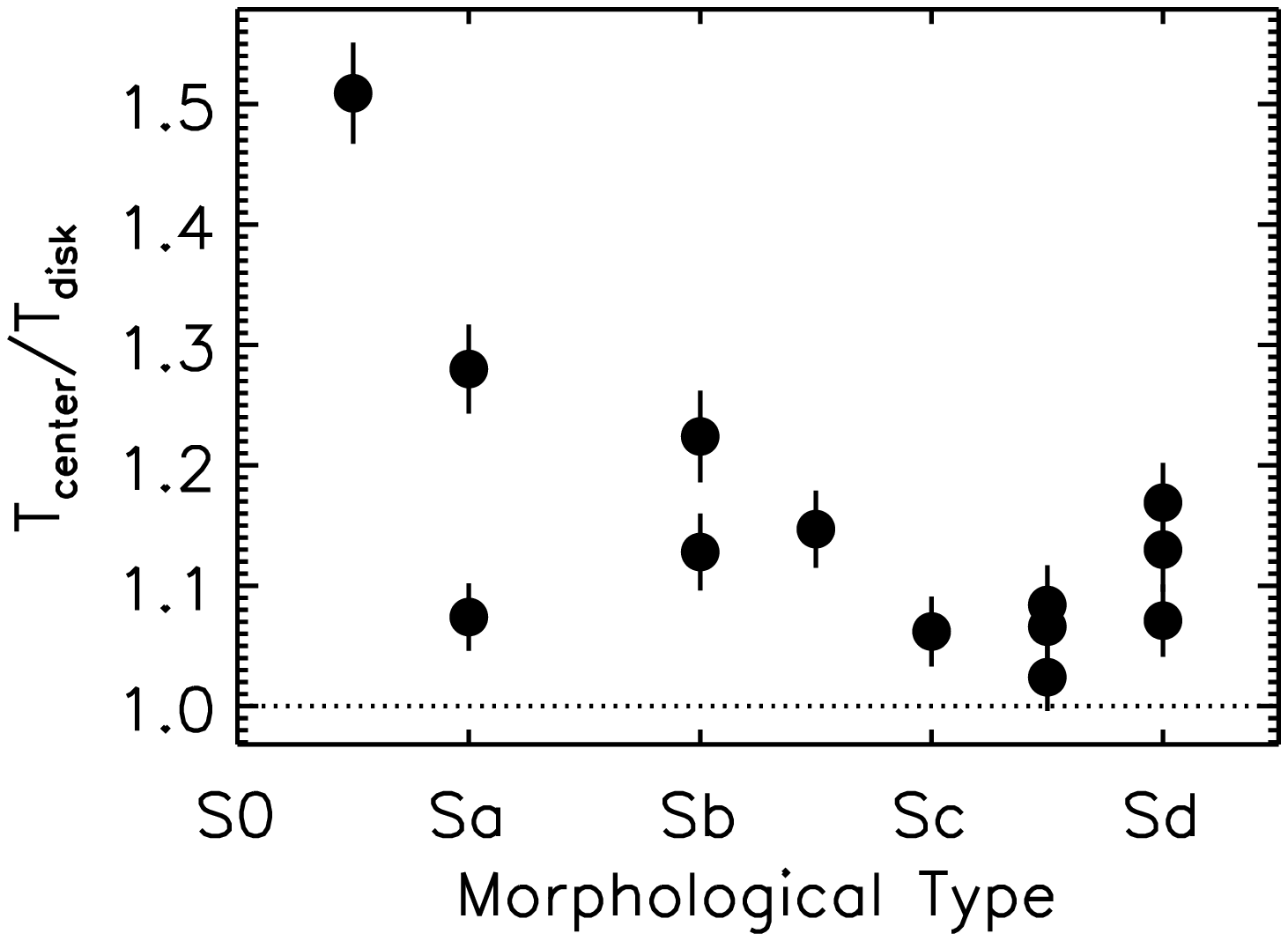}
\caption{Ratio of central dust temperature to disk dust temperature as a
function of morphological type.  The dotted line is drawn at a ratio of 1
(i.e., at equal temperatures) to guide the eye.\label{fig:ratios}}
\end{figure}

\begin{figure}
\centering
\includegraphics[width=0.49\textwidth]{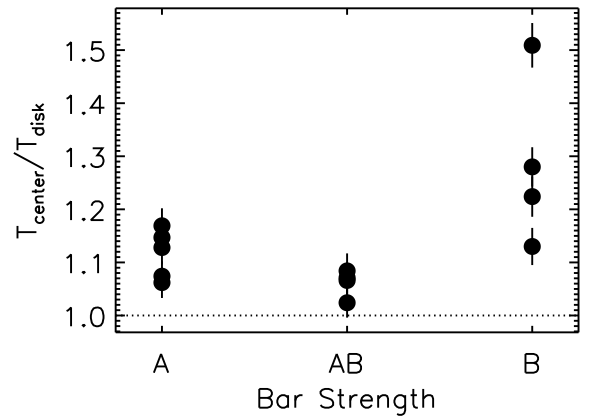}
\caption{Ratio of central dust temperature to disk dust temperature as a
function of bar strength, as indicated by morphological type.  The dotted line
is drawn at a ratio of 1 (i.e., at equal temperatures) to guide the
eye.\label{fig:ratios_v_bar}}
\end{figure}

For each galaxy, we computed a ratio of the central to disk temperature (from
data at wavelengths longer than 70~$\mu$m) of the cool dust component.  The
average ratio of central-to-disk temperatures in this sample is $1.15 \pm
0.03$.  We have plotted this ratio against morphological type in
Figure~\ref{fig:ratios}, where the average ratio is $1.23 \pm 0.03$ for types
earlier than Sc and $1.09 \pm 0.03$ for later types.  The trend is poorly
described by a line, because it has a correlation coefficient of 0.66.  We
plot the ratio against bar strength (as defined by the morphological type) in
Fig.~\ref{fig:ratios_v_bar}, where the average ratio is $1.29 \pm 0.04$ for
strong bars and $1.09 \pm 0.03$ for weak bars.

A significant ($3\sigma$) correlation between temperature ratio and Hubble
type is observed in Fig.~\ref{fig:ratios}, although the correlation is only
marginal, at $2\sigma$, if the most significant point (NGC~1291) is removed
from the correlation.  This trend suggests that the dominant stellar bulges in
early-type spirals are able to heat the cool dust in the ISM to greater
temperatures than their late-type counterparts.  The effect is most pronounced
in the earliest type galaxy in our sample, NGC~1291, in which the dust in the
central region is almost 50\% warmer than the dust in the disk.  In comparing
this to the spectral types in Table~\ref{tab:sample}, we have found no obvious
correlation between the temperature ratio and the presence of a central
nonthermal source in our galaxies.

Bars may play an important role in driving the correlation of temperature
ratio with morphology.  As shown in Fig.~\ref{fig:ratios_v_bar}, a strong
bar is correlated with a high center/disk temperature ratio, although the
scatter in the strong-bar ratios is large.  This could occur through
bar-induced starbursts, since mid-infrared colors are redder in barred
galaxies, implying warmer dust temperatures (Huang et~al.  \cite{huang96};
Roussel et~al.\ \cite{roussel01}).  Barred galaxies also tend to have higher
star-formation rates (SFRs) than unbarred galaxies, possibly because bars
drive gas into the inner regions where it can pile up in the inner dynamical
resonances associated with rings (Hawarden et~al.\ \cite{hawarden86}).  The
bar-driven gas transport has been shown in CO observations (e.g., Sakamoto
et al.\ \cite{sakamoto99}).  Stars will form quite readily in such conditions,
where the gas transported by the bar fuels star formation (Erwin \& Sparke
\cite{erwin02}; B{\"o}ker et~al.\ \cite{boker08}; Comer{\'o}n et~al.\
\cite{comeron08}).  In fact, the frequency of inner rings in early-type barred
spirals (Hunt \& Malkan \cite{hunt99}; Erwin \& Sparke \cite{erwin02}) could
conspire to produce the effect seen in our sample.  Similarly, bars are more
frequently found in bulge-dominated galaxies (e.g., Masters et~al.\
\cite{masters10}), which would also contribute to the trends we see.  The
concentrations of CO gas in the centers have been observed in early-type
spiral galaxies with intense star formation (Koda et al.\ \cite{koda05}).
In addition to a dependence on stellar density, the higher temperature
ratios in the early types shown here could therefore be associated with bar-induced star
formation.  This is consistent with the result we get if the 70~$\mu$m fluxes
are not included in the modified blackbody fits.  Without them, the correlation
between temperature ratio and either Hubble type or bar strength disappears;
i.e., measurements at wavelengths shorter than the peak emission are required
to determine the dust temperature accurately.

\section{Conclusion}

We used far-infrared data from \emph{Spitzer} and submillimeter data from
\emph{Herschel} to compute separate SEDs for the center and disk regions of 13
nearby galaxies observed as part of the KINGFISH program.  We fit those SEDs
(at wavelengths longer than 70~$\mu$m) with blackbody functions (modified by a
frequency-dependent emissivity) to compute temperatures.  On average, the cool
dust temperature of the central component is $15 \pm 3$\% hotter than the
disk.  We find that the central temperature is higher than the disk by 20\% to
50\% in galaxies of type S0 to Sb, but only 9\% higher in later types.  This
ratio is also higher (at $1.29 \pm 0.04$) in strongly barred galaxies than in
weakly barred galaxies (at $1.09 \pm 0.03$).

The data therefore indicate that the large (or ``classical'') grains that
dominate the far-infrared and submillimeter emission are warmer in the centers
of those galaxies with a substantial bulge and/or a strong bar.  This may simply be
caused by the higher density of the radiation field in the centers of early-type
spirals, enhanced star formation due to the bar, or some combination of the
two.  A cleaner separation of morphological components (perhaps with larger
samples and/or less distant galaxies) and a more thorough assessment of the
density of starlight and star formation activity, plus an evaluation of the
impact of central nonthermal sources, may help separate these effects.

The analysis presented here illustrates the power of {\it Herschel}
observations in characterizating the spatially resolved distribution of
dust in nearby galaxies.  This power will grow with the use of better-resolved
far-infrared SEDs as measured by PACS (Poglitsch et al.\ \cite{poglitsch10}),
which will let us measure smaller and/or more distant galaxies and
determine radial trends of dust temperature.

\begin{acknowledgements}
The following institutes have provided hardware and software elements to the
SPIRE project:  University of Lethbridge, Canada; NAOC, Beijing, China;  CEA
Saclay, CEA Grenoble, and OAMP in France;  IFSI, Rome, and University of Padua,
Italy; IAC, Tenerife, Spain; Stockholm Observatory, Sweden; Cardiff
University, Imperial College London, UCL-MSSL, STFC-RAL, UK ATC Edinburgh, and
the University of Sussex in the UK.  Funding for SPIRE has been provided by
the national agencies of the participating countries and by internal institute
funding: CSA in Canada; NAOC in China; CNES, CNRS, and CEA in France; ASI in
Italy; MCINN in Spain; Stockholm Observatory in Sweden; STFC in the UK; and
NASA in the USA.  Additional funding support for some instrument activities
has been provided by ESA.  We would also like to thank the anonymous
referee whose comments helped improve this paper.
\end{acknowledgements}

\end{document}